\documentclass[12pt]{article}
\usepackage{latexsym,epsfig,graphicx,amsmath,amssymb,amscd}
\usepackage{eucal,undertilde,multirow,paralist}
\usepackage[american]{babel}
\usepackage{epstopdf}
\usepackage{amsthm}
\usepackage[nodayofweek]{datetime}
\usepackage[round]{natbib}
\usepackage{xr-hyper}

\usepackage[bookmarks,bookmarksnumbered=true,colorlinks=true,
     linkcolor=blue,urlcolor=blue,citecolor=blue,breaklinks = true,backref=page]{hyperref}

\input{commands.sty}

\newcommand{\domainfun}{{\mathbb{D}}}

\newdateformat{mydate}{\twodigit{18}{  }March 2024}

\newcommand{\footremember}[2]{%
	\footnote{#2}
	\newcounter{#1}
	\setcounter{#1}{\value{footnote}}%
}

\title{On Equivalence of Likelihood-Based Confidence Bands for Fatigue-Life and Fatigue-Strength Distributions}
\author{
  Peng Liu \footremember{jmp}{JMP Statistical Discovery LLC}
  \and Yili Hong\footremember{vt}{Department of Statistics, Virginia Tech}
  \and Luis A. Escobar\footremember{lsu}{Department of Experimental Statistics, Louisiana State University}%
  \and William Q. Meeker\footremember{isu}{Department of Statistics, Iowa State University}
}
\mydate
\begin{document}
\maketitle

\begin{abstract}
Fatigue data arise in many research and applied
areas and there have been statistical
methods developed to model and analyze such data.
The distributions of fatigue life and fatigue strength
are often of interest to engineers designing products that might
fail due to fatigue from cyclic-stress loading. Based on a specified statistical
model and the maximum likelihood method, the cumulative distribution
function (cdf) and quantile function (qf) can be estimated for the
fatigue-life and fatigue-strength distributions.
Likelihood-based confidence bands
then can be obtained for the cdf and qf.  This
paper provides equivalence results for confidence bands for
fatigue-life and fatigue-strength models. These results
are useful for data analysis and computing
implementation. We show (a) the
equivalence of the confidence bands for the fatigue-life cdf
and the fatigue-life qf,
(b) the equivalence of confidence bands for the fatigue-strength cdf and
the fatigue-strength qf, and (c) the
equivalence of  confidence bands for the fatigue-life
qf and the
fatigue-strength qf. Then we
illustrate the usefulness of those equivalence results with two examples
using experimental fatigue data.
\end{abstract}


\mydate
\tableofcontents


\ifempty
\section*{Acronyms}
\begin{center}
\begin{tabular}{lll}
cdf & \quad &cumulative distribution function\\
pdf & \quad &probability density function\\
qf & \quad &quantile function\\
\SN{}  & \quad &Stress versus number of cycles
\end{tabular}
\end{center}

\section*{Notation}
\begin{center}
\begin{tabular}{cll}
$N$ &\quad & cycles to failure\\
$N_{e}$ &\quad & specific level of the number of cycles to failure where
  estimation of fatigue-strength characteristics are needed\\
$S$ & \quad & level of stress (or strain)\\
$S_{e}$ & \quad & specific level of stress (or strain) where
  estimation of fatigue-life characteristics are needed\\
$\epsilon$ & \quad & random variable that has a standard
  location-scale distributions and that is pare of the random-error
  term of a statistical model
  distribution\\
$\sigma_{N}$ & \quad & shape parameter of a specified log-location-scale
  fatigue-life distribution\\
$\sigma_{X}$ & \quad & shape parameter of a specified log-location-scale
  fatigue-strength distribution\\
\end{tabular}
\end{center}
\fi

\section{Introduction}
\subsection{Background}

Fatigue-life data from cyclic-stress experiments arise in many
research and development areas. Fatigue-life experiments are usually
conducted by allocating a sample of test specimens to different
fixed levels of stress or strain amplitude (depending on material
properties and the degree of loading some test are conducted by
controlling strain and others are conducted by controlling stress or
some stress-related variable like displacement). Units are run until
failure (defined in some precise way such as crack initiation,
specimen fracture, or noticeable delamination) or until a specified
test-end time. Units that reach the end of the test without failing
are right-censored (also known as runouts).

Traditional methods of modeling experimental fatigue data
typically focused modeling to estimate the
characteristics (such as quantiles and tail probabilities)
of the fatigue-life distribution. Engineers, however, are often
also interested in the quantiles or tail probabilities of the
the fatigue-strength distribution. Based on a specified statistical
model, the maximum likelihood (ML) method can be used to estimate the
cumulative distribution function (cdf) and the quantile function
(qf) for both the fatigue-life and fatigue-strength distributions.
Then, likelihood-ratio-based confidence intervals (LR confidence
intervals) can be obtained for
quantiles and cdf probabilities of these distributions.

In this paper, we prove three important equivalence results
that are important in applications:
\begin{itemize}
\item
The equivalence, for a given level of stress $S_{e}$, of LR confidence bands for the fatigue-life cdf and qf.
\item
The equivalence, for a given level of cycles $N_{e}$, of LR confidence bands for the fatigue-strength cdf and
  fatigue-strength qf.
\item The equivalence, for a given probability level $p$, of LR confidence bands for the fatigue-life qf
  and fatigue-strength qf.
\end{itemize}
These three equivalence results are not well-known,
but are useful in data analysis and computing implementations, as
demonstrated in our numerical examples.

\subsection{Fatigue-Life and Fatigue-Strength Random Variables}
\label{section:fatigue.life.and.fatigue.strength}

This section briefly describes the relationship between
the fatigue-life and fatigue-strength random variables.
The distributions of these two closely-related
random variables are often needed to quantify the reliability of a
product or system component that can fail from fatigue.
Technical details describing this
relationship are given in \citet{Meekeretal2022}
and summarized later in this paper.

\begin{figure}
\begin{tabular}{cc}
Lognormal distributions~(a) & Weibull distributions~(b) \\[-3.2ex]
\rsplidapdffiguresize{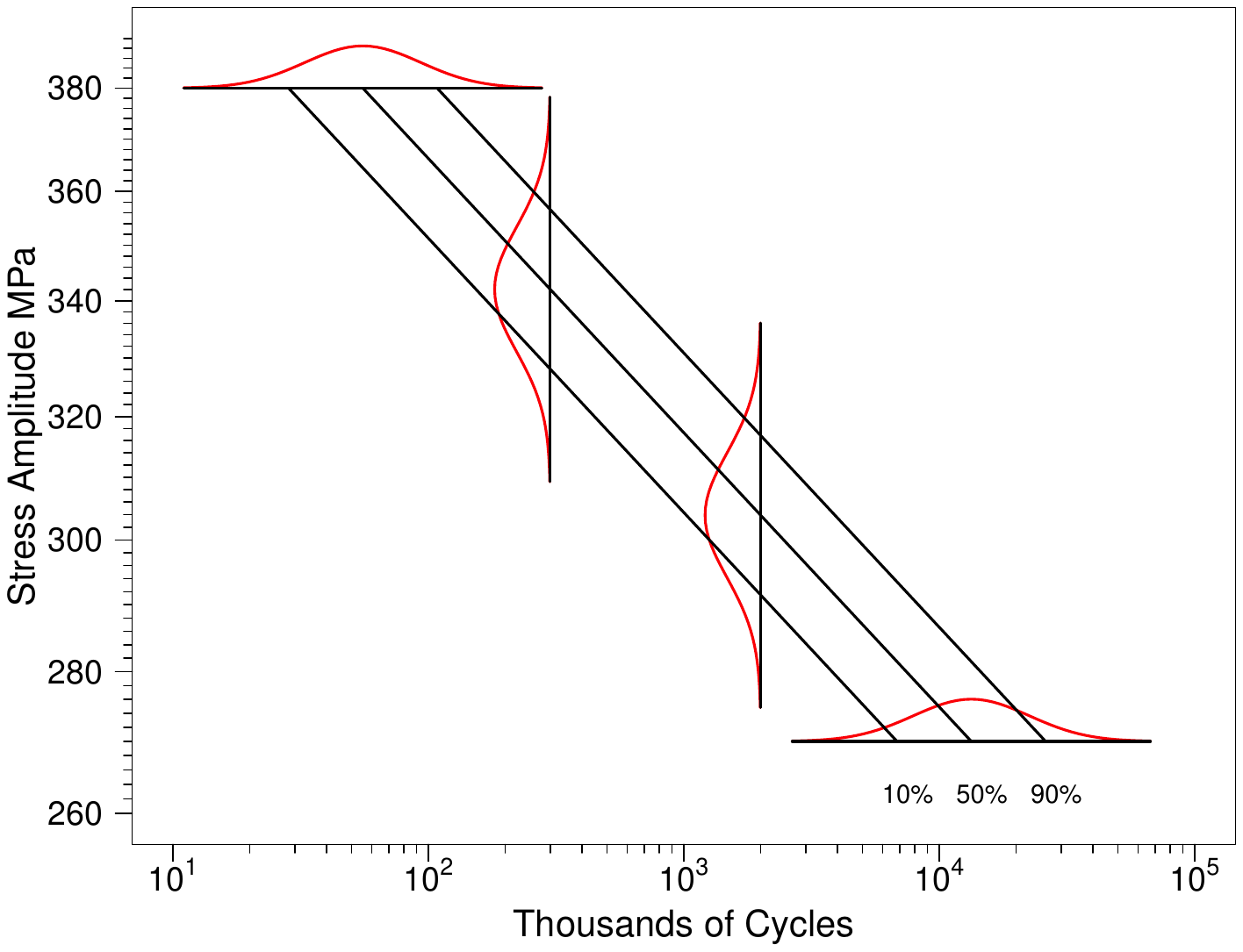}{3.2in}&
\rsplidapdffiguresize{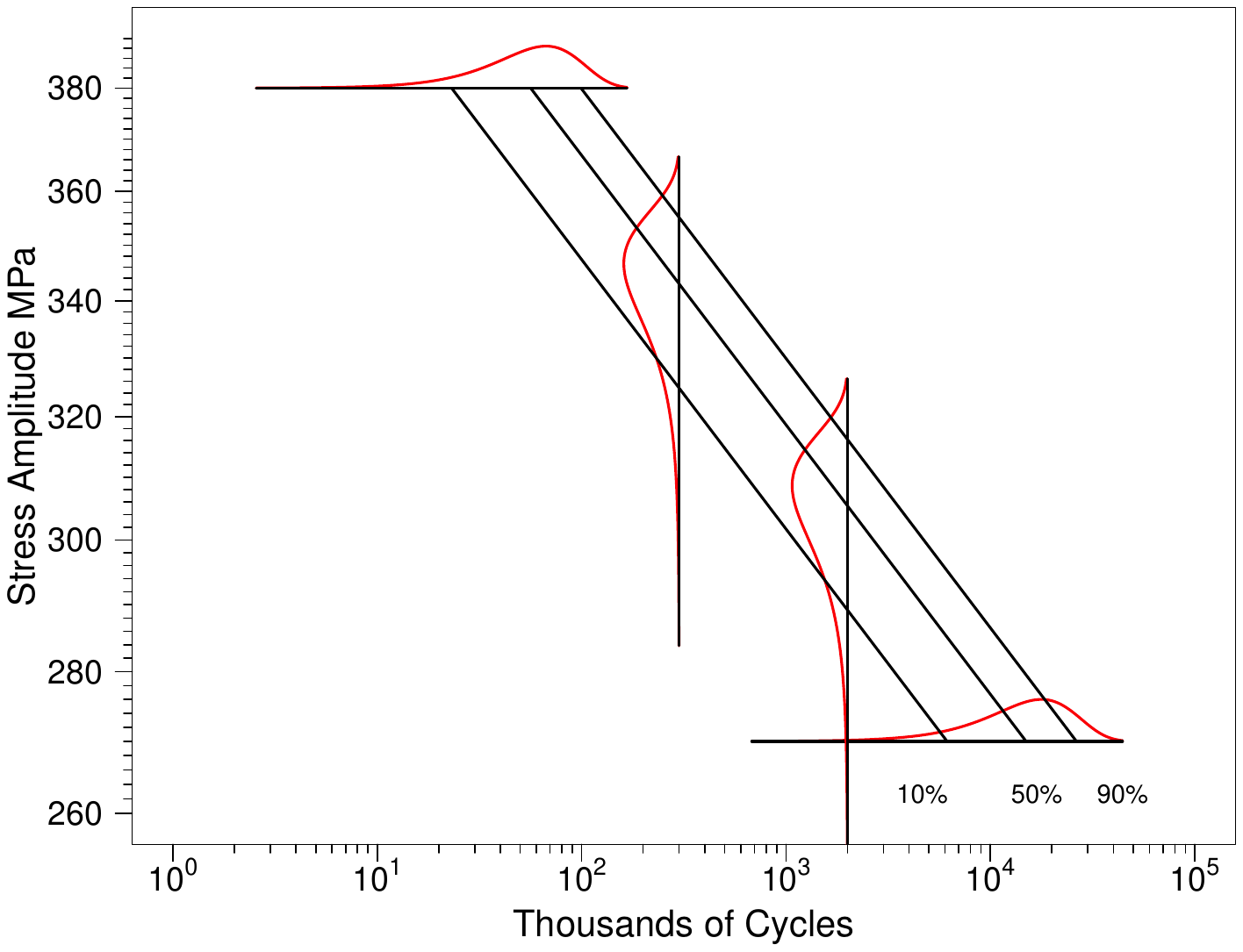}{3.2in}\\
\end{tabular}
\caption{Plot showing fatigue-life and fatigue-strength lognormal distributions~(a); Plot showing fatigue-life and fatigue-strength Weibull distributions~(b).}
\label{figure:fatigue.life.fatigue.strength.plots}
\end{figure}

Fatigue life $N$ is the number of cycles when a unit fails from cyclic
loading. The cdf for $N$ depends on stress (or strain) amplitude, $S$. For some
experiments, there may be other experimental variables (e.g., the
stress ratio, testing temperature, and dwell time). In this paper we
consider fatigue-life distributions that depend on stress
amplitude. Extensions are possible (and straightforward),
as described in the example in
Section~\ref{section:estimating.quantiles.fatigue.strength.distribution.new.spring}.
The horizontal densities in
Figure~\ref{figure:fatigue.life.fatigue.strength.plots}a are
lognormal fatigue-life distributions and those in
Figure~\ref{figure:fatigue.life.fatigue.strength.plots}b are Weibull
fatigue-life distributions on the right. The Weibull and lognormal
are the most frequently used probability distributions for
fatigue-life. For the examples in
Figure~\ref{figure:fatigue.life.fatigue.strength.plots}, the scale
parameters of the distributions depend on the level of stress and
the shape parameters do not.

The random variable $X$, fatigue-strength, is the
\textit{level of stress} at which a unit would fail at a given
number of cycles, $N_{e}$.  The
vertical densities in
Figure~\ref{figure:fatigue.life.fatigue.strength.plots}a are
lognormal fatigue-strength distributions  and  those in
Figure~\ref{figure:fatigue.life.fatigue.strength.plots}b are Weibull
fatigue-strength distributions.
Fatigue strength cannot be observed directly
because the stress levels in the test, $S$, are specified experimental levels
and the failure time (the number of cycles, $N$) is random. However, the
fatigue-strength distribution can be estimated from fatigue-life data.

As illustrated in Figure~\ref{figure:fatigue.life.fatigue.strength.plots},
the fatigue-life and fatigue-strength models share the same quantile
lines. The relationships between the models can also be expressed
in a precise manner, as they are in
\citet[][Sections~2 and~3]{Meekeretal2022}.
A summary of those results is presented in
Sections~\ref{sec:confidence.bands.induced.fatigue.strength.distribution}
and~\ref{section:induced.fatigue.life.model.and.ml}.

Traditionally, fatigue-life data are analyzed by specifying a model
for the fatigue-life random variable $N$, resulting in an induced
model for the fatigue-strength $X$. Another approach is to specify
the fatigue-strength model, resulting in an induced model for
fatigue-life. These approaches are equivalent only if
the \SN{} relationship is a straight line (as it is in
Figure~\ref{figure:fatigue.life.fatigue.strength.plots}) when
plotted on log-log scales.  There are
important advantages of latter approach
when the \SN{} relationship is nonlinear. These advantages are
described in \citet[][Section~3.2.1]{Meekeretal2022}. We discuss the
this new approach in
Section~\ref{section:statistical.inference.specified.fatigue.strength.model}
and illustrate its use in the corresponding example in
Section~\ref{section:estimating.quantiles.fatigue.life.nitinol.wire}.

\subsection{Related Literature and Contributions of This Work}

For a single distribution,
\citet{HongMeekerEscobar2008b} showed that
pointwise likelihood-ratio-based confidence interval bands
for a cdf are exactly the same as pointwise likelihood-ratio-based
confidence interval bands for the corresponding qf.
\citet{Meekeretal2022} provide a comprehensive
review of the traditional modeling approach for fatigue-life data and
present a novel alternative modeling approach using modern
statistical methods for modeling fatigue life as a function of
applied cyclic stress amplitude. Their results can be used to
estimate both a fatigue-life distribution
at a given level of stress and a fatigue-strength distribution
at a given
level of cycles. The contribution of this paper
is to extend the results in  \citet{HongMeekerEscobar2008b}
to linear or nonlinear
regression models that connect fatigue-life models and
fatigue-strength models and to show how existing software
for estimating
fatigue-life models can be used to obtain confidence intervals
for fatigue-strength quantities of interest. We prove three
important equivalence results and show how
they provide insights that are useful
for data analysis and computing implementation.

\subsection{Overview}

The rest of this paper is organized as
follows. Section~\ref{sec:data.model} introduces the data and model
and shows how to obtain pointwise LR confidence
intervals and confidence bands for
a \textit{specified} fatigue-life model
and for the corresponding \textit{induced}
fatigue-strength model.
Section~\ref{section:equivalence.results.specified.fatigue.life.model}
shows, for a specified fatigue-life model, the equivalence of
confidence bands for
(a) the fatigue-life cdf and fatigue-life qf
(b) the fatigue-strength cdf and
fatigue-strength qf,
and (c) the fatigue-strength qf
and fatigue-life qf.
Section~\ref{section:statistical.inference.specified.fatigue.strength.model}
describes similar results for the important situation
when the fatigue-strength model is specified and the fatigue-life
model is induced.
Numerical examples illustrating the usefulness
of our results for the two different modeling
approaches are given in
Sections~\ref{section:estimating.quantiles.fatigue.strength.distribution.new.spring}
and~\ref{section:estimating.quantiles.fatigue.life.nitinol.wire}, respectively.
Section~\ref{section:conclusions}
contains some conclusions and areas for future research.

\section{Data, Model, and Statistical Inference with a Specified
  Fatigue-Life Model}
\label{sec:data.model}
\subsection{Data and Model}
This section introduces notation and the
traditional statistical model for describing the fatigue-life data. The \SN{} data
(stress/strain and number of cycles to failure),
often with right censoring, are denoted by $(S_{i}, N_{i},
\delta_{i})$, $i=1, \dots, n$. Here $S_{i}$ is the level of
applied stress (or strain) amplitude (which,
for simplicity of expression,
we refer to as ``stress''), $N_{i}$ is the observed cycles
to failure or number of cycles at the end of the test
for a right-censored observation
(called a runout in the fatigue literature),
and $\delta_{i}$ is the failure indicator, where
\begin{align*}
\delta_{i}&=
\begin{cases}
1 & \text{if $N_{i}$ is a failure time} \\
0 & \text{if $N_{i}$ is a runout time.} \\
\end{cases}
\end{align*}

Let $N$ be the number of cycles to failure and let $S$ be the
applied stress amplitude. The fatigue-life model can be written as
\begin{align}
\label{S.equation:fl.const.sigma.model}
\log(N)&=\log[g(S;\betavec)]+\sigma_{N}\epsilon,
\end{align}
where $g(S;\betavec)$ is a positive monotonically
decreasing (often nonlinear) function of stress $S$ with
regression parameter vector $\betavec$. The random error term
is $\sigma_{N}\epsilon$ where $\epsilon$
has a location-scale distribution with $\mu=0$ and $\sigma=1$ and
$\sigma_{N}$ is a shape parameter for the log-location-scale distribution
describing the variability in  $N$.
Then $\thetavec=(\betavec', \sigma_{N})'$ is a vector
containing the unknown regression-model parameters.
The lognormal and Weibull distributions are the most commonly used
distributions for fatigue life and both are members of the
log-location-scale family of distributions, as described in
\citet[][Chapter~4]{MeekerEscobarPascual2021}.

To simplify the presentation in this section we will suppose
that $\log[g(S;\betavec)]$ has neither a vertical nor a horizontal
asymptote. When such asymptotes exist, the results are similar but
require some technical adjustments \citep[described in Section~2.4
of][]{Meekeretal2022}.

Let $F_{N}(t ;S, \thetavec)$ denote the fatigue-life cdf and let
$f_{N}(t; S, \thetavec)=d\,F_{N}(t; S, \thetavec)/dt$
denote the corresponding fatigue-life pdf. Then for any given stress level
$S_{e}$, fatigue life $N$ has a log-location-scale distribution with
cdf
\begin{align}
\label{equation:fatigue.life.failure.time.model.cdf}
F_{N}(t; S_{e}, \thetavec)&=\Pr \left(N \le t; S_{e}\right)=\Phi \left
(\frac{\log(t)-\log[\gfun(S_{e};\betavec)]}{\sigma_{N}} \right),
\quad \quad t>0, \,\, S_{e}>0,
\end{align}
where $\gfun(S_{e};\betavec)$ is a scale parameter and $\sigma_{N}$
is a shape parameter of the distribution of $N$.
For any given stress level
$S_{e}$, the fatigue-life
$p$ quantile is
\begin{align}
\label{equation:fatigue.life.failure.time.model.quantile}
t_{p}(S_{e}, \thetavec) &= \exp[\log[\gfun(S_{e};\betavec)] +
\Phi^{-1}(p)\sigma_{N}], \quad \quad 0<p<1, \,\, S_{e}>0.
\end{align}
To simplify the presentation, we will often omit the dependency on
$\thetavec$ and write $t_{p}(S_{e})$.  When we refer to
$t_{p}(S_{e})$ as a ``quantile function'' (qf) it could be (for fixed
$\thetavec$) $t_{p}(S_{e})$ as a function of $p$ for fixed $S_{e}$
or $t_{p}(S_{e})$ as a function of $S_{e}$ for fixed $p$. The exact
meaning will be clear from the context of the use. We will use
similar notation and terminology for a fatigue-strength qf.

\subsection{Maximum Likelihood Estimation}
\label{section:maximum.likelihood.estimation}
The log-likelihood function for
the fatigue \SN{} data is
\begin{align}
\label{equation:location.scale.likelihood}
\loglike(\thetavec) &= \sum_{i=1}^{n}\left\{
\delta_{i} \log  \left[f_{N}(N_{i} ; S_{i}, \thetavec)
\right ] +
(1-\delta_{i}) \log \left [1- F_{N}(N_{i} ;S_{i}, \thetavec) \right]
\right \}.
\end{align}
The ML estimator $\thetavechat$ is
the value of $\thetavec$ that
maximizes~(\ref{equation:location.scale.likelihood}).
Standard optimization algorithms can be used to maximize
$\loglike(\thetavec)$.

For our goal of fatigue-life experiments, we are interested in
estimators of quantiles and probabilities of the fatigue-life and/or
fatigue-strength distributions. Based on the invariance property of
maximum likelihood estimators, the maximum likelihood estimators of 
$t_{p}(S_{e}, \thetavec)$, $F_{N}(t;S_{e},\thetavec)$,
$x_{p}(N_{e}, \thetavec)$, and $F_{X}(x;N_{e},\thetavec)$ 
are respectively
$t_{p}(S_{e}, \thetavechat)$, $F_{N}(t;S_{e},\thetavechat)$,
$x_{p}(N_{e}, \thetavechat)$, and $F_{X}(x;N_{e},\thetavechat)$.
In the next section, we describe the procedure to calculate
likelihood-ratio-based confidence intervals  (LR intervals) for these quantities,
which are all scalar functions of parameters.

\subsection{Likelihood-Ratio-Based Confidence Intervals}
This section describes procedures to calculate
LR confidence intervals
for a scalar function of parameters.
The first procedure is based on the profile relative likelihood
function which provides a useful visualization of likelihood
inference for a scalar function of the model parameters. The second
procedure provides the same intervals but has important
computational and technical advantages that we use in establishing
the main results of this paper.

\subsubsection{Likelihood-ratio-based confidence intervals for a single parameter}
\label{sec:likelihood.ci.single.parameter}
The procedure for computing LR intervals for
a scalar parameter is well-known. For example, to calculate an
interval for $\theta_{i}$, element $i$ of the full
parameter vector $\thetavec$,
construct a profile relative likelihood function for
$\theta_{i}$ which is
\begin{align}
\label{equation:profile.likelihood.single.parameter}
R(\theta_{i}) &= \max_{\thetavec \backslash \{\theta_{i}\}}\left[\frac{\exp[\loglike(\thetavec)]}{\exp[\loglike(\thetavechat)]}\right],
\end{align}
where $\thetavec \backslash \{\theta_{i}\}$ denotes the parameter
vector $\thetavec$ with $\theta_{i}$ excluded (i.e., the nuisance
parameters).  The profile relative likelihood function
is described and illustrated in
\citet[][Section~8.2]{MeekerEscobarPascual2021}.  An approximate
$100(1-\alpha)\%$ LR confidence interval is defined by
the set of all values of $\theta_{i}$ where $R(\theta_{i})$ exceeds
$\exp[-(1/2)\chi_{(1-\alpha; 1)}^2]$, where $\chi_{(p; 1)}^2$ is the
$p$ quantile of a chi-square distribution with $1$ degree of
freedom. That this is so can be seen by noting that these values
of $\theta_{i}$ are the values that would not be rejected by a
likelihood-ratio test.

The endpoints of the LR confidence interval can
written as $\left[\,\undertilde{\theta_{i}},\quad
  \tildex{\theta_{i}} \,\right]$ and can be defined more formally as
\begin{align}
  \begin{split}
\label{equation:theta.i.profile.interval}
\undertilde{\theta_{i}}&=\min\big\{ 
	\theta_{i} \mid R(\theta_{i})
	 \geq \exp(-(1/2)\chi_{(1-\alpha; 1)}^2)
\big\}\\ 
\tildex{\theta_{i}}&=\max\big\{ 
	\theta_{i} \mid R(\theta_{i})
	 \geq \exp(-(1/2)\chi_{(1-\alpha; 1)}^2) 
\big\},
    \end{split}
\end{align}
where the notation $\big\{ e \mid \mathbb{C} \big\}$ is the set for
which every element $e$ satisfies the condition $\mathbb{C}$. A
numerical implementation of this approach requires finding the roots
of the profile relative likelihood function. In this paper we assume
that the likelihood is concave. 
This will be so if the statistical model is
identifiable and there is a sufficient amount of data.
Under this condition, the relative likelihood $R(\theta_{i})$ is unimodal 
and any interval generated by the procedure will be closed.
Meanwhile, the roots $\undertilde{\theta_{i}}$ and $\tildex{\theta_{i}}$,
where $R(\theta_{i})$ intersects $\exp(-(1/2)\chi_{(1-\alpha; 1)}^2)$
may not exist, in which case the corresponding confidence interval 
endpoint is on or approaching the boundary of the parameter space.
That, however, does not affect subsequent results.

\subsubsection{Likelihood-ratio-based confidence intervals for a scalar function
of the parameters}
\label{section:likelihood.ci.scalar.function.parameters}

To find an LR confidence interval for a scalar function
$\xi(\thetavec)$ of the parameters, the same procedure can be used after a
reparameterization where $\xi(\thetavec)$ replaces any one of the
elements of $\thetavec$. This can be done because of the invariance
properties of likelihood inference.  Such reparameterizations are
described and illustrated
\citet[][Section~8.2.4]{MeekerEscobarPascual2021} and
are easy to implement if it is easy to compute $\xi(\thetavec)$ and
its inverse. When there is no closed form for the inverse (a
frequently occurring situation), numerical
inversion can be used but that requires much more computing effort.

When pedagogy is less important, there are computational and
technical reasons to use the 
log-likelihood (instead of the relative likelihood) directly to find LR intervals.
Using the log-likelihood directly, the same interval endpoints in
(\ref{equation:theta.i.profile.interval})
can be obtained from:
\begin{align*}
\undertilde{\theta_{i}}&=\min\big\{ 
	\theta_{i} \mid
	\loglike(\thetavec) = k 
\big\}\\ 
\tildex{\theta_{i}}&=\max\big\{ 
	\theta_{i} \mid 
	\loglike(\thetavec) = k 
\big\},
\end{align*}
where $k=\loglike(\thetavechat)-(1/2)\chi_{(1-\alpha; 1)}^2$. This approach
recognizes that to define the interval,
it is only necessary to find the interval endpoints 
and suggests a completely different formulation. The
operation of maximizing out the nuisance parameters is replaced by
the constraint $\loglike(\thetavec) = k$ because all candidate
interval endpoints have to be on a particular
log-likelihood level curve. The indicated minimization and maximization are
done subject to that constraint. More importantly, this constrained
optimization approach greatly simplifies the procedure for finding
LR confidence intervals for functions of the parameters
because reparameterization is not needed. That is the LR confidence
interval $\left[\,\undertilde{\xi}(\thetavec),\quad
  \tildex{\xi}(\thetavec) \,\right]$
for $\xi(\thetavec)$ can be obtained from
\begin{align}
  \begin{split}
\label{equation:xi.maxL}
\undertilde{\xi}(\thetavec)&=\min\big\{ 
	\xi(\thetavec) \mid \loglike(\thetavec) = k 
\big\}\\ 
\tildex{\xi}(\thetavec)&=\max\big\{ 
	\xi(\thetavec) \mid \loglike(\thetavec) = k 
\big\}.
    \end{split}
\end{align}
Standard optimization algorithms (e.g., the Lagrange
multiplier method) can be used to do the needed constrained optimization.
\citet{Doganaksoy2021} uses this approach, to compute LR confidence
intervals in several applications where reparameterization would be difficult.
His presentation contains details of the needed optimization algorithms.

Similar to~(\ref{equation:theta.i.profile.interval}), 
in this paper we assume that the part of $\xi(\thetavec)$ 
subject to the constraint $\loglike(\thetavec) = k$ is continuous.
This will be so if $\xi(\thetavec)$ is a continuous function of $\thetavec$.
Meanwhile, the endpoints $\undertilde{\xi}(\thetavec)$ 
and $\tildex{\xi}(\thetavec)$ subject to the constraint $\loglike(\thetavec) = k$
may not exist, in which case the corresponding confidence interval 
endpoint is on or approaching the boundary of the parameter space.
That, however, does not affect subsequent results.

\subsection{Likelihood-Ratio-Based Confidence Bands}
\label{sec:confidence.bands}

Sets of pointwise confidence intervals are
generally computed and displayed in statistical software along with
point estimates of qfs and cdfs.  We call these sets of intervals
``pointwise confidence bands.'' Often Wald-based intervals are used
to compute the pointwise confidence bands
in these applications because they require less computational
effort. LR bands, however, have desirable technical properties
(e.g., coverage probabilities that are closer to the nominal
confidence level and the equivalence results given in this paper);
such LR bands are defined in this section.

\subsubsection{Likelihood confidence bands for fatigue-life
  qfs and cdfs when the
  fatigue-life model is specified}
\label{sec:confidence.bands.specified.fatigue.life.distribution}
For any given stress level
$S_{e}$,  the $p$ quantile of $N$ is
$t_{p}(S_{e}) = \exp[\log[\gfun(S_{e};\betavec)] + \Phi^{-1}(p)\sigma_{N}],$
which is a function of $\thetavec$,
 as in~(\ref{equation:fatigue.life.failure.time.model.quantile}).
An LR $100(1-\alpha)\%$ 
confidence interval for the fatigue-life cdf $t_{p}(S_{e})$
for given $S_{e}$ and value of $p$
using~(\ref{equation:xi.maxL}) is
$\left[\,\tplower(S_{e}),\quad\tpupper(S_{e}) \,\right]$
where
\begin{align}
  \begin{split}
\label{S.equation:lik.ci.tp}
\tplower(S_{e})&=\min\big\{ t_{p}(S_{e},\thetavec ) \mid \loglike(\thetavec) = k
\big\}\\ 
\tpupper(S_{e})&=\max\big\{ t_{p}(S_{e}, \thetavec) \mid \loglike(\thetavec) = k
\big\},
    \end{split}
\end{align}
$k=\loglike(\thetavechat )-(1/2)\chi_{(1-\alpha; 1)}^2$, and
$\thetavechat = (\betavechat',\sigmahat_{N})'$.
Again, standard
optimization algorithms can be used to do the needed optimization.
Note that when used for a given value of $S_{e}$ and range of values of $p$,
(\ref{S.equation:lik.ci.tp}) provides a set of pointwise
confidence intervals for the fatigue-life qf
$t_{p}(S_{e})$ which we refer to
as a ``pointwise confidence band'' for the qf.

Similarly, an LR  $100(1-\alpha)\%$ confidence interval for
the fatigue-life cdf $F_{N}(t; S_{e})$, a function of $\thetavec$,
at given stress level $S_{e}$ and time $t$ is
$[\,\, \Flower_{N}(t; S_{e}),\quad \Fupper_{N}(t; S_{e}) \,\,]$,
where
\begin{align}
  \begin{split}
\label{S.equation:prob.lik.ci}
\Flower_{N}(t; S_{e})&=\min\big\{F_{N}(t; S_{e}, \thetavec) \mid \loglike(\thetavec) = k
\big\}\\ 
\Fupper_{N}(t; S_{e})&=\max\big\{F_{N}(t; S_{e}, \thetavec) \mid \loglike(\thetavec) = k
\big\}.
    \end{split}
\end{align}
Again, when used over a given range of $t$ values
(\ref{S.equation:prob.lik.ci}) provides a pointwise confidence band
for $F_{N}(t; S_{e})$, the fatigue-life cdf at stress
level $S_{e}$ for that range of $t$ values.

\subsubsection{Likelihood confidence bands for an induced fatigue-strength
 qf and an induced fatigue-strength
  cdf  when the fatigue-life model is given}
\label{sec:confidence.bands.induced.fatigue.strength.distribution}
As described in Section~\ref{section:fatigue.life.and.fatigue.strength},
fatigue strength is an unobservable random variable $X$ defined as the
level of stress where failure would occur at $N_{e}$, a given
number of cycles. As shown in \citet[][Section~2.4.1]{Meekeretal2022},
to obtain the fatigue-strength cdf at $N_{e}$,
start with (\ref{S.equation:fl.const.sigma.model}) and replace $N$
by $N_{e}$ and $S_{e}$ by $X$ giving
\begin{align*}
\log(N_{e})&=\log[g(X;\betavec)] + \sigma_{N}\epsilon.
\end{align*}
Thus the random variable driven by $\sigma_{N}\epsilon$ is switched from
$N$ at the given value of $S_{e}$ to $X$ at given value of $N_{e}$. Then
solving for the fatigue-strength random variable $X$ gives
\begin{align*}
X &= g^{-1}[\exp(\log(N_{e})-\sigma_{N}\epsilon); \betavec].
\end{align*}
The cdf of fatigue strength $X$ is
\begin{align*}
F_{X}(x; N_{e}) & =\Pr(X\leq x)=\Pr(\log[g(X;\betavec)]\geq
\log[g(x;\betavec)])\\ & =\Pr(\log(N_{e})-\sigma_{N}\epsilon \geq
\log[g(x;\betavec)])\\ & =\Pr\left[\epsilon \leq \frac{\log(N_{e})-
    \log[g(x;\betavec)]}{\sigma_{N}}\right]\\ &
=\Phi\left[\frac{\log(N_{e})-
    \log[g(x;\betavec)]}{\sigma_{N}}\right].
\end{align*}
This cdf is \textit{not} a log-location-scale distribution unless,
for fixed $\betavec$, $\log[g(x;\betavec)])$ is a linear function
of $\log(x)$.

The qf for $X$ is
$x_{p}(N_{e})=g^{-1}[\exp[\log(N_{e})-\Phi^{-1}(p)\sigma_{N} ];\betavec].$
Then, using~(\ref{equation:xi.maxL}), for given values of cycles
$N_{e}$ and $p$, the $100(1-\alpha)\%$ LR
confidence interval for $x_{p}(N_{e})$, a function of $\thetavec$, is
$\left[\,\xplower(N_{e}),\quad\xpupper(N_{e})\,\right]$
where
\begin{align}
  \begin{split}
\label{S.equation:lik.ci.xp}
\xplower(N_{e})
&=\min\big\{ x_{p}(N_{e},\thetavec) \mid \loglike(\thetavec) = k
                      \big\}\\
\xpupper(N_{e})
&=\max\big\{ x_{p}(N_{e},\thetavec) \mid \loglike(\thetavec) = k
                      \big\}.
    \end{split}
\end{align}
Note that (\ref{S.equation:lik.ci.xp}) also provides pointwise
confidence bands for $x_{p}(N_{e})$, the fatigue-strength
qf at $N_{e}$ cycles for a given range
of values of $p$.

Similarly, the LR confidence interval for fatigue-strength cdf
$F_{X}(x; N_{e})$, a function of $\thetavec$, 
for a given value $N_{e}$ cycles at a given $x$ is
$[\,\, \Flower_{X}(x; N_{e}),\quad \Fupper_{X}(x; N_{e}) \,\,]$
where
\begin{align}
  \begin{split}
\label{S.equation:prob.lik.ci.xe}
\Flower_{X}(x; N_{e}) &=\min\big\{F_{X}(x; N_{e}, \thetavec) \mid \loglike(\thetavec) = k
\big\}\\ 
\Fupper_{X}(x; N_{e}) &=\max\big\{F_{X}(x; N_{e}, \thetavec) \mid \loglike(\thetavec) = k
\big\}.
    \end{split}
\end{align}
Thus, (\ref{S.equation:prob.lik.ci.xe})
provides pointwise confidence bands for $F_{X}(x; N_{e})$, the
fatigue-strength cdf for a given range of $x$ values.

\section{Equivalence Results when  the Fatigue-Life Model is Specified}
\label{section:equivalence.results.specified.fatigue.life.model}
This section shows the important equivalence results for
LR confidence bands. Note that for confidence bands
computed in other ways (e.g., the simpler and widely used Wald method),
the results will be approximate.

\subsection{General Equivalence Result for Two Mutually Inverse Monotone Functions}
\label{sec:equivalence.results.inverse.functions}
This section presents a general result that can then be used to
easily establish the particular results that are important in practical
applications.

\begin{result}
\label{S.inverse.function.result:eqv}
The pointwise LR confidence bands for
two strictly monotone (increasing or decreasing) functions 
that are inverses of each other are equivalent.
\end{result}

\begin{figure}
\begin{tabular}{cc}
(a)~LR confidence bands of~$\xi(w)$ & (b)~Transposed bands of~$\xi^{-1}(v)$ \\[-2.5ex]
\rsplidapdffiguresize{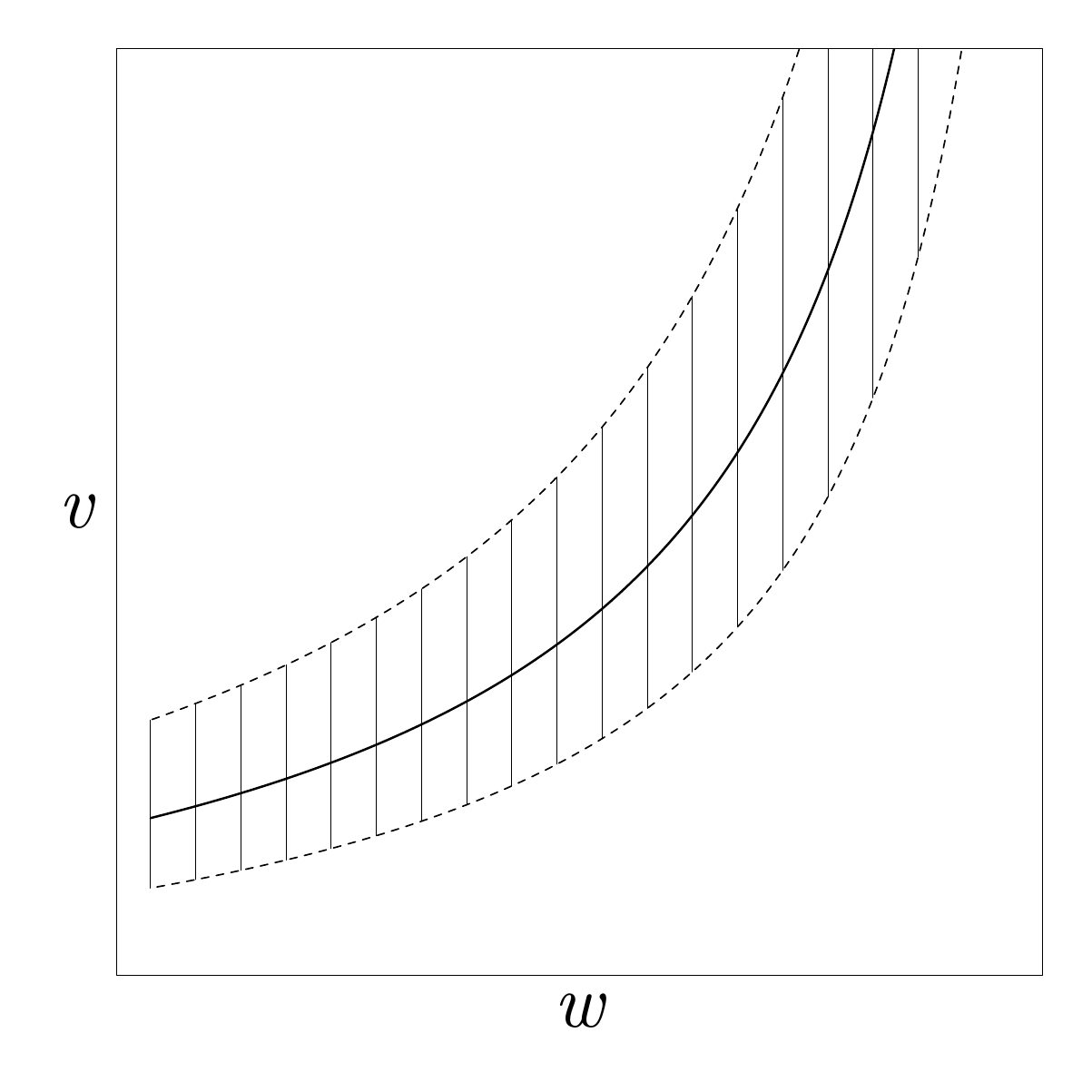}{3.2in}&
\rsplidapdffiguresize{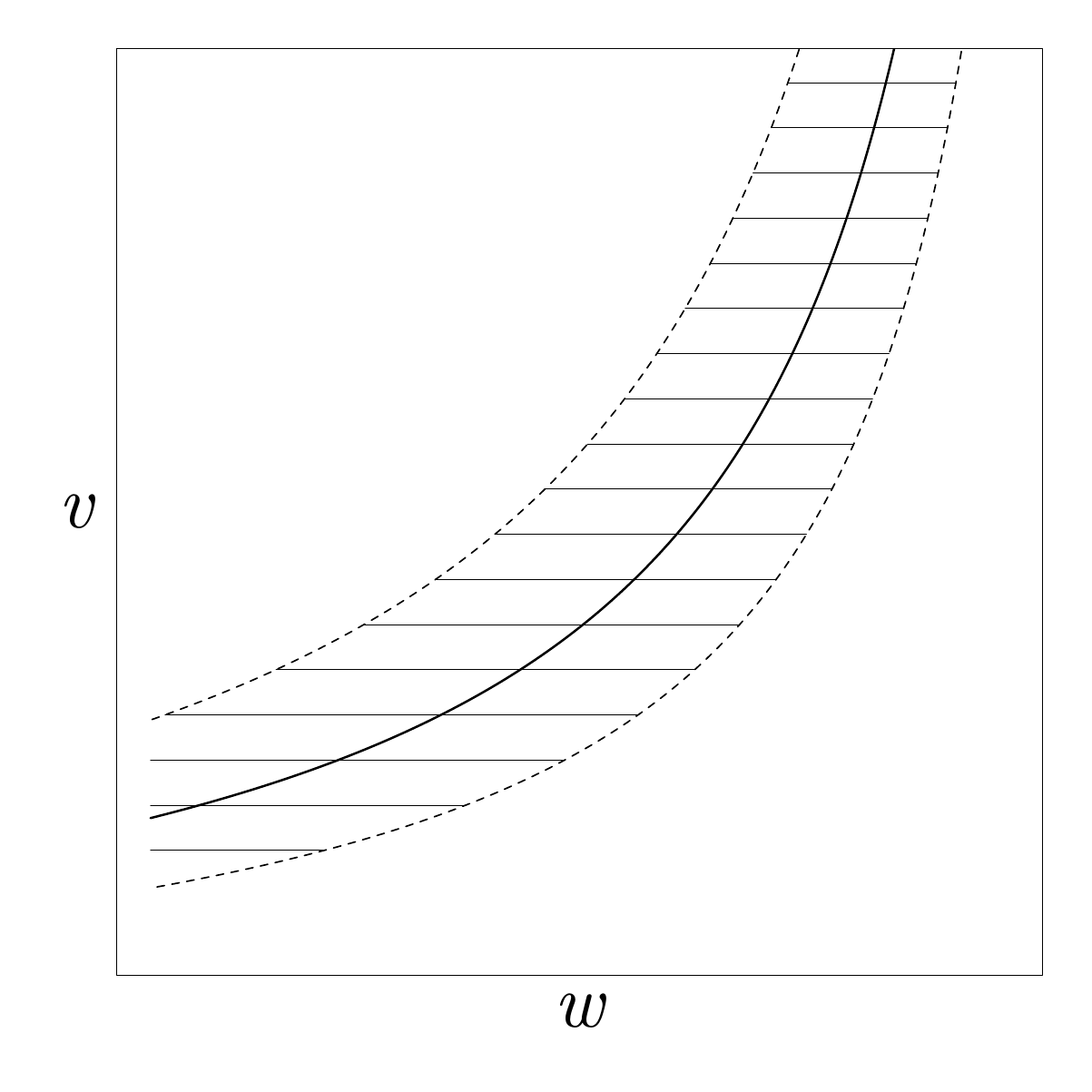}{3.2in}\\
\end{tabular}
\caption{Plot showing LR confidence bands for  $\xi(w)$ as a collection of individual confidence intervals of $\xi(w)$ at individual $w$~(a); 
Plot showing a transposed LR confidence bands for  $\xi^{-1}(v)$ as a collection of individual confidence intervals of $\xi^{-1}(v)$ at individual $v$~(b).}
\label{figure:ci.bands.illustration}
\end{figure}

To show the result, we first extend (\ref{equation:xi.maxL}) as
follows
\begin{align}
\label{equation:interval.equal.point.set}
\left\{ \left[\,\undertilde{\xi}(\thetavec),\quad
  \tildex{\xi}(\thetavec) \,\right] \mid \loglike(\thetavec) = k 
\right\} &\equiv
\big\{ 
	\xi(\thetavec) \mid \loglike(\thetavec) = k 
\big\},
\end{align}
which states the equivalence of two sets.
The set on the left contains a single interval. The set on the right
contains an infinite number of individual points.
The sets are equivalent in the sense that the interval 
in the set on the left
encloses all points in the set on the right.

Next, suppose we have a strictly monotone function
$v=\xi(w,\thetavec)$, whose inverse function is
$w=\xi^{-1}(v,\thetavec)$. By being strictly monotone,
we ensure that for any $w$ there is one and only one
corresponding $v$, and vice versa.
The $100(1-\alpha)\%$
pointwise LR confidence bands for $\xi(w, \thetavec)$,
using~(\ref{equation:interval.equal.point.set}),
can be defined in two equivalent ways
\begin{align}
\label{S.equation:vfunction.ci.region1}
B_{\xi(w)}&=\left\{
	\left[   \undertilde{\xi}(w, \thetavec),\quad
	\tildex{\xi}(w, \thetavec) \right]
 \mid
	\forall w \in \domainfun_{w}, 
	\loglike(\thetavec) = k 
\right\}\\
\label{S.equation:vfunction.ci.region2}
&\equiv \big\{
	(w,v) \mid 
	\forall w \in \domainfun_{w}, 
	\forall v\in \domainfun_{v}, 
	v=\xi(w, \thetavec), 
	\loglike(\thetavec) = k
\big\},
\end{align} 
where $\domainfun_{w}$ denotes the domain of
$\xi(w, \thetavec)$ (and the range of $\xi^{-1}(v, \thetavec)$).
Also, $\domainfun_{v}$ denotes the domain of
$\xi^{-1}(v, \thetavec)$ (and the range of $\xi(w, \thetavec)$).
The first definition~(\ref{S.equation:vfunction.ci.region1})
is a set of confidence intervals for the function
$\xi(w, \thetavec)$  for $w$ spanning the domain of the function,
illustrated by Figure~\ref{figure:ci.bands.illustration}a.
There are an infinite number of intervals in this set.
According to~(\ref{equation:interval.equal.point.set}),
each single interval set 
$\left\{\left[   \undertilde{\xi}(w, \thetavec),\quad
	\tildex{\xi}(w, \thetavec) \right]
\mid
w \in \domainfun_{w},
\loglike(\thetavec) = k
\right\}$ 
for a given $w$
can be expressed equivalently as a set of points $\big\{v
\mid
w \in \domainfun_{w},
v = \xi(w, \thetavec), \loglike(\thetavec) = k
\big\}$ for the same $w$.
By collecting the equivalent sets of points for all $w$,
the second definition~(\ref{S.equation:vfunction.ci.region2})
expresses the set of intervals 
equivalently as a set 
of points $(w,v)$ that are constrained by
the functional relationship
$v=\xi_{v}(w, \thetavec)$ and $\loglike(\thetavec) = k$.

For the inverse function,
the $100(1-\alpha)\%$
pointwise LR confidence bands for $\xi^{-1}(v, \thetavec)$,
using~(\ref{equation:interval.equal.point.set}),
can be defined in two equivalent ways as well:
\begin{align}
\label{S.equation:ufunction.ci.region1}
B_{\xi^{-1}(v)}&=\left\{
	\left[   \undertilde{\xi^{-1}}(v, \thetavec),\quad
	\tildex{\xi^{-1}}(v, \thetavec) \right]
\mid
	\forall v \in \domainfun_{v}, 
	\loglike(\thetavec) = k
\right\}\\
\label{S.equation:ufunction.ci.region2}
&\equiv \big\{
	(v,w) \mid
	\forall w\in \domainfun_{w}, 
	\forall v \in \domainfun_{v}, 
	w = \xi^{-1}(v, \thetavec), 
	\loglike(\thetavec) = k
\big\} \\
\label{S.equation:ufunction.ci.region3}
&=\big\{
	(v,w) \mid
	\forall w\in \domainfun_{w}, 
	\forall v \in \domainfun_{v}, 
	v = \xi(w, \thetavec), 
	\loglike(\thetavec) = k
\big\}.
\end{align} 
The first definition~(\ref{S.equation:ufunction.ci.region1}) 
is a set of individual confidence intervals
for function
$\xi^{-1}(v, \thetavec)$ for $v$ spanning the domain of the function,
illustrated by Figure~\ref{figure:ci.bands.illustration}b.
There are an infinite number of intervals in this set.
The intervals in Figure~\ref{figure:ci.bands.illustration}b 
have been transposed to illustrate the inverse relationship.
The second definition~(\ref{S.equation:ufunction.ci.region2})
expresses the set of intervals equivalently
as a set of all points
$(v,w)$  that are constrained by
the functional relationship
$w = \xi^{-1}(v, \thetavec)$ and $\loglike(\thetavec) = k$.
The sets~(\ref{S.equation:ufunction.ci.region1})  
and~(\ref{S.equation:ufunction.ci.region2}) are equivalent
because of the similar reason 
for the equivalence between~(\ref{S.equation:vfunction.ci.region1})
and~(\ref{S.equation:vfunction.ci.region2}).
The sets~(\ref{S.equation:ufunction.ci.region2})
and~(\ref{S.equation:ufunction.ci.region3}) are equal
because we just replace the condition
$w = \xi^{-1}(v, \thetavec)$ with the equivalent
condition $v = \xi(w, \thetavec)$.

By comparing the sets~(\ref{S.equation:vfunction.ci.region2})
and~(\ref{S.equation:ufunction.ci.region3}), we have
$\forall (v,w)\in B_{\xi(w)}$, $(w,v) \in B_{\xi^{-1}(v)}$, and vice versa.
In other words, the transpose of $B_{\xi(w)}$ is identical to $B_{\xi^{-1}(v)}$,
and vice versa. 
Thus the transposed pointwise confidence bands for $v=\xi(w,\thetavec)$
are exactly the same as the pointwise confidence bands for  $w=\xi^{-1}(v,\thetavec)$.

It is interesting to note that the result holds even if
the function has horizontal or vertical asymptotes,
as long as the function is strictly monotone.
For example, suppose $\xi(w,\thetavechat)$ is the 
center curve in Figure~\ref{figure:ci.bands.illustration}a,
which is the maximum likelihood estimate of $\xi(w,\thetavec)$.
Suppose the function $\xi(w, \thetavechat)$
has a vertical asymptote at $w_{0}$ on the right side. 
Then $\xi(w, \thetavechat)$ is undefined on $[w_{0}, \infty)$.
The most important question, however, is whether 
the likelihood based pointwise confidence intervals of $\xi(w, \thetavec)$
are defined on $[w_{0}, \infty)$. Without the loss of 
generality, we just need to check whether the interval for
$\xi(w_{0}, \thetavec)$ is defined, which is, according
to~(\ref{S.equation:vfunction.ci.region1}),
$\left\{
	\left[   \undertilde{\xi}(w_{0}, \thetavec),\quad
	\tildex{\xi}(w_{0}, \thetavec) \right]
 \mid
	\loglike(\thetavec) = k 
\right\}$, or equivalently,
according to~(\ref{S.equation:vfunction.ci.region2}),
$\big\{
	(w_{0},v) \mid 
	\forall v\in \domainfun_{v}, 
	v=\xi(w_{0}, \thetavec), 
	\loglike(\thetavec) = k
\big\}$. Whether the length of the interval is zero,
or whether the set of points is empty, those situations
do not affect the above result, which depends
 only on the conditions of how the sets are constructed.

This concludes the demonstration that the
pointwise LR confidence bands are equivalent for
two monotone functions that are inverse to each other.
With this general result, we can immediately obtain
the next two results.

\subsection{Equivalence Results for cdf and qf Bands}
\label{sec:equivalence.results.cdf.quantile.pairs}

The following result extends a similar result in
\citet{HongMeekerEscobar2008b} to the linear and nonlinear
regression models used here.
\begin{result}
\label{S.result:eqv}
The LR confidence bands for 
the specified fatigue-life cdf in 
(\ref{S.equation:prob.lik.ci}) 
and 
the corresponding fatigue-life qf in 
(\ref{S.equation:lik.ci.tp}) are equivalent.
\end{result}
The result holds as an application of
Result~\ref{S.inverse.function.result:eqv}
because the fatigue-life cdf
and the fatigue-life quantiles are
inverse functions of each other.
This result shows that if one uses the LR procedures,
it makes no difference whether one computes pointwise confidence
bands for $F_N(t; S_e)$ or $t_{p}(S_{e})$---the bands will be the
same.  

The following shows that the equivalence result extends to the
induced fatigue-strength model.
\begin{result}
\label{S.Strength.result:eqv}
The LR confidence bands
for the induced
fatigue-strength cdf in
(\ref{S.equation:lik.ci.xp})
and the corresponding qf in  (\ref{S.equation:prob.lik.ci.xe})
are equivalent.
\end{result}
The result holds as an application of
Result~\ref{S.inverse.function.result:eqv}
because the induced fatigue-strength cdf and
and the corresponding fatigue-strength quantiles are
inverse functions of each other.
This result shows that if one uses the LR procedures,
it makes no difference whether one computes pointwise confidence
bands for $x_{p}(N_{e})$ or $F_X(x; N_e)$---the bands will be the
same.

\subsection{Equivalence of Confidence Bands for the Fatigue-Life
  qf and the Induced Fatigue-Strength qf}
\label{sec:Equivalence.fatigue.life.fatigue.strength.quantile.confidence.bands}

Consider the need to find a safe level of stress $S_{e}$ such that
$\tplower(S_{e})=N_{e}$. If computed using the LR
method, this $S_{e}$ value is equivalent to the lower confidence
bound $\xplower(N_{e})$ of the fatigue-strength distribution. More
generally, if pointwise LR confidence intervals for
$t_{p}(S_{e})=N_{e}$ are used to obtain bands of such $S_{e}$ values
for a range of $N_{e}$ values, they will be equivalent to the
confidence bands for $x_{p}(N_{e})$ in (\ref{S.equation:lik.ci.xp})
over the same range of $N_{e}$. We have the following result.

\begin{result}
\label{S.Strength.lift:eqv} For a fixed $p$, the $p$ qf curve for
the fatigue-life distribution as a function of $S_{e}$ is the
same as the $p$ qf curve for the fatigue-strength distribution as a
function of $N_{e}$. In addition, for fixed $p$, the confidence
bands for the fatigue-life
qf in \eqref{S.equation:lik.ci.tp} over a range of $S_{e}$ and the
induced fatigue-strength qf in \eqref{S.equation:lik.ci.xp} over a
range of $N_{e}$ are equivalent.
\end{result}

We first show that the $p$ qf curve for the fatigue-life
distribution is the same as the $p$ qf curve for the
fatigue-strength distribution.  For a fixed $p$ and $S_e$, a point on
the qf for the fatigue-life distribution is $(S_e,
\exp[\mu_{\betavec}+\Phi^{-1}(p)\sigma_N ])$, where
$\mu_{\betavec}=\log[g(S_{e};\betavec)]$.
This point is also on the $p$ qf for the
fatigue-strength distribution for a fixed
$N_e=\exp(\mu_{\betavec}+\Phi^{-1}(p)\sigma_N)$.
The qf for $X$ is
\begin{align*}
x_p(N_e)&=g^{-1}[\exp(\log(N_e)-\Phi^{-1}(p)\sigma_N );\betavec]\\
&=g^{-1}[\exp(\log[\exp(\mu_{\betavec}+\Phi^{-1}(p)\sigma_N )]-\Phi^{-1}(p)\sigma_N );\betavec]\\
&=g^{-1}[\exp(\mu_{\betavec});\betavec]=S_e,
\end{align*}
which is the same point as $(S_e, \exp(\mu_{\betavec}+\Phi^{-1}(p))\sigma_N)$.
Thus the two qf curves are equivalent.

Because the two qf curves are equivalent, the two different
quantiles
functions are inverses of each other. Then, as an application of
Result~\ref{S.inverse.function.result:eqv},
the confidence bands for the fatigue-life
qf in \eqref{S.equation:lik.ci.tp} and the induced fatigue-strength qf in
\eqref{S.equation:lik.ci.xp} are then equivalent.

\section{Estimating Quantiles of the Fatigue-Strength
  Distribution for a New Spring}
\label{section:estimating.quantiles.fatigue.strength.distribution.new.spring}
A large experiment was conducted to estimate the fatigue-life
distribution of a newly designed nib spring to be used in a
vehicle. In a full-factorial experiment, nine springs were tested at
each combination of two levels of a manufacturing-processing
temperature (500 and 1{,}000 degrees F), two manufacturing methods
(Old and New), and three levels of Stroke (50, 60, and 70 mils), for
a total of 108 springs. Each spring was tested until failure or
5{,}000 thousand cycles (whichever came first). Stroke, proportional
to stress, was the displacement used in the cyclic stressing of the
springs and served as an accelerating variable. That is, the
engineers were interested in the reliability of the springs at lower
levels of Stroke that would be encountered in actual vehicle
operation. There were 73 observed failure times and 35 runouts by
the end of the test. The data were presented and previously analyzed in
\citet{MeekerEscobar2003} and
\citet[][Chapter~19]{MeekerEscobarPascual2021}. Details of the ML
estimation will not be repeated here.

Although the question was not explicitly addressed in these
references, the engineers who requested the data analysis wanted to
know the level of Stroke  such that the spring
would survive 500,000 thousand cycles (500 million cycles) with only
a 0.10 probability of failing. In other words, the engineers wanted
a lower confidence bound on the 0.10 quantile of the
fatigue-strength distribution.

\begin{figure}
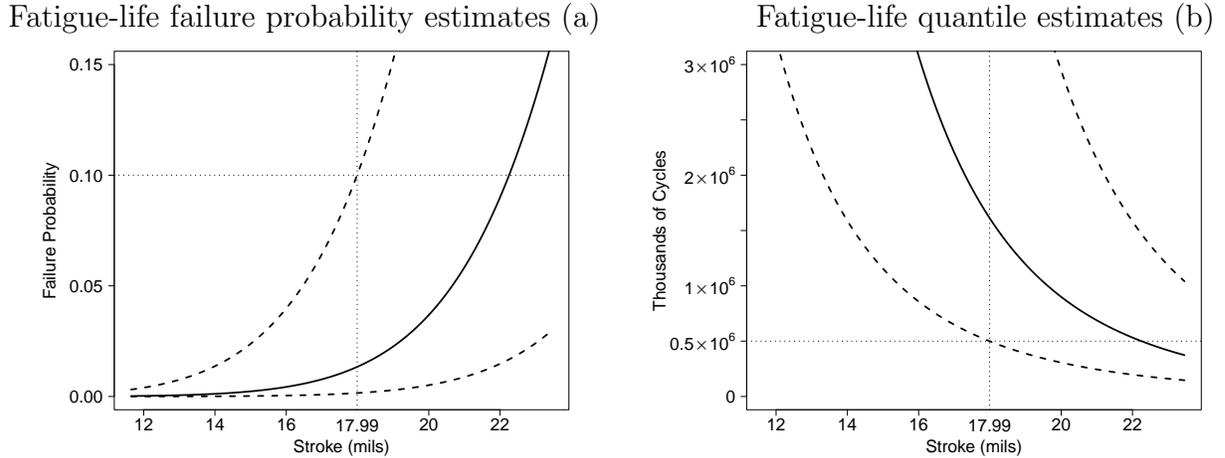

\begin{tabular}{cc}
\phantom{x}Fatigue-life failure probability estimates~(a) & \phantom{xxxxxxx}Fatigue-life quantile estimates~(b) \\[-2ex]
\rsplidapdffiguresize{pdfFigures/Figure2a_NewSpringProbabilityProfilerLikelihoodIntervals}{3.1in}&
\rsplidapdffiguresize{pdfFigures/Figure2b_NewSpringQuantileProfilerLikelihoodIntervals}{3.1in}\\
\end{tabular}
\caption{Fatigue-life cdf estimates at 500{,}000 cycles as a
  function of Stroke~(a); Fatigue-life 0.10 quantile estimates as a
  function of Stroke~(b); the dashed lines are pointwise 90\%
  confidence bands.}
\label{figure:newspring.fatigue.life.plots}
\end{figure}

Figure~\ref{figure:newspring.fatigue.life.plots}a is a plot of the
ML estimates of the fatigue-life failure probabilities
and corresponding
pointwise LR two-sided 90\% confidence bands, as a
function of Stroke for the New manufacturing method and 600 degrees
F temperature. Using
Result~\ref{S.result:eqv} and Result~\ref{S.Strength.result:eqv} in
Section~\ref{sec:equivalence.results.cdf.quantile.pairs},
Figure~\ref{figure:newspring.fatigue.life.plots}a
shows that the LR one-sided lower
95\% confidence bound for the desired 0.10 quantile of the
fatigue-strength distribution at 500{,}000 cycles
can be obtained from the value of
Stroke that results in a one-sided \textit{upper} 95\% confidence bound
$\tildex{F}_{N}(500{,}000; \textrm{Stroke})=0.10$.
The figure (note where the dotted lines
cross) shows that this value is $\undertilde{x}_{0.10}(500{,}000) = 17.99$ mils.

Figure~\ref{figure:newspring.fatigue.life.plots}b shows the ML
estimates and LR two-sided 90\% confidence intervals
for the fatigue-life 0.10 quantile as a function of Stroke. This
plot illustrates an alternative way to obtain the desired lower
confidence bound on the 0.10 quantile of the fatigue-strength
distribution by using Result~\ref{S.Strength.lift:eqv} in
Section~\ref{sec:Equivalence.fatigue.life.fatigue.strength.quantile.confidence.bands}. In
particular, one finds the level of Stroke that has a \textit{lower}
confidence bound for the fatigue-life
quantile equal to $\undertilde{t}_{0.10}(\textrm{Stroke})=500{,}000$
cycles (where the dotted lines
cross), again giving $\undertilde{x}_{0.10}(500{,}000) = 17.99$ mils.

Plots like those
in Figure~\ref{figure:newspring.fatigue.life.plots} can be obtained from
commonly available statistical
software (e.g., JMP, Minitab, or SAS) for fitting
lifetime regression models (although Wald-based intervals are
usually provided, at least as a default).
This example, using our results, shows how
existing inferential methods for fatigue-life probabilities and quantiles
can be used to make inferences on fatigue-strength quantiles and probabilities.

\section{Statistical Inference with a Specified
  Fatigue-Strength Model}
\label{section:statistical.inference.specified.fatigue.strength.model}
In contrast to the traditional fatigue-life modeling methods
described in Section~\ref{sec:data.model} (that are currently widely
available in statistical software), this section describes
the alternative approach where the fatigue-strength model is
specified and the fatigue-life model is induced (and these methods
have not been available is widely available statistical software).

\subsection{The Induced Fatigue-Life Model and Maximum Likelihood
  Estimation}
\label{section:induced.fatigue.life.model.and.ml}
This section briefly reviews the approach described in
\citet[][Section~3.2]{Meekeretal2022} in which one specifies a
fatigue-strength model which induces a fatigue-life model that is
fit to the fatigue-life data.  An important advantage of this
approach is that, as pointed out by \citet{Weibull1956}
\citep[and also mentioned in][]{BastenaireBastienPomey1961}, the
shape/spread of fatigue-strength distributions tend not to depend on
$N_{e}$, the given number of cycles. This is in contrast to the
shape/spread of fatigue-life distributions that often do depend on
$S_{e}$, the given level of stress or strain (especially in
high-cycle fatigue applications).
\citet[][Section~3.2.1]{Meekeretal2022}
provides further discussion about this point.

Recall (from Section~\ref{section:fatigue.life.and.fatigue.strength})
that fatigue strength is defined to be the level of stress
amplitude that will cause a unit to fail at a given number of
cycles $N_{e}$. This random variable is not observable but its
distribution can be estimated by using appropriate
\SN{} data.  Let $X$
be the fatigue strength and let $N_{e}$ be the given number of
cycles to failure. A model for fatigue-strength can be written as
\begin{align}
\label{equation:specified.fatigue.strength.model}
\log(X)&=\log[h(N_{e};\betavec)]+\sigma_{X}\epsilon,
\end{align}
where $h(t;\betavec)$ is a positive monotonically decreasing
function of $t$ for a fixed regression parameter vector
$\betavec$. This model implies that for fixed values of $N_{e}$ and
$\betavec$, $X$ has a log-locations-scale distribution with a scale
parameter $h(N_{e};\betavec)$. The random error term is
$\sigma_{X}\epsilon$ where $\epsilon$ has a location-scale
distribution with $\mu=0$ and $\sigma=1$ and $\sigma_{X}$ is a shape
parameter for the log-location-scale distribution describing the
variability in fatigue strength $X$.

As in Section~\ref{sec:data.model}, to simplify the presentation in this section
we will also suppose that $\log[h(t;\betavec)]$ has neither a
vertical nor a horizontal asymptote. When such asymptotes exist, the
results are similar but require some technical adjustments
\citep[described in Section~3.2~of~][]{Meekeretal2022}.

Omitting the details of the derivation \citep[which are given in
Section~3.2.3 of][]{Meekeretal2022}, after specifying a model for
fatigue strength $X$, one can obtain, for a given level of applied
stress amplitude $S_{e}$, the cdf for the fatigue-life random
variable $N$ as
\begin{align}
\label{equation:induced.fatigue.life}
F_{N}(t; S_{e})&=\Pr(N \le t)=\Phi\left[\frac{\log(S_{e})-\log[\hfun(t;\betavec)]}{\sigma_{X}}   \right], \quad \quad
  t>0, \,\, S_{e}>0.
\end{align}
This cdf is \textit{not} a log-location-scale distribution unless,
for fixed $\betavec$, $\log[\hfun(t;\betavec)]$ is a linear function
of $\log(t)$.

Because $N$ is observable,
the cdf in (\ref{equation:induced.fatigue.life}) along with the
corresponding pdf $f_{N}(t; S_{e})= dF_{N}(t; S_{e})/dt$ can be used
to define the likelihood for the available fatigue-life data and to
obtain ML estimates of the parameter vector $\thetavec=(\betavec',
\sigma_{X})'$, as described in
Section~\ref{section:maximum.likelihood.estimation}. Sections~\ref{section:likelihood.conf.bands.specified.fatigue.strength.distribution}
and~\ref{section:likelihood.conf.bands.induced.fatigue.life.distribution}
show how to compute LR confidence intervals/bands, for
cdfs and qfs for the fatigue-strength distributions
and induced fatigue-life distributions, respectively.

\subsection{Likelihood Confidence Bands for the Specified Fatigue-Strength Distribution}
\label{section:likelihood.conf.bands.specified.fatigue.strength.distribution}
Let $\loglike\left(\betavec,\sigma_{N}\right)$ be the log-likelihood
function, and $\betavechat$ and $\sigmahat_{N}$ be the ML estimates.
For any given number of cycles $N_{e}$,
\eqref{equation:specified.fatigue.strength.model}
implies that $X$ has a log-location-scale distribution with log-location
parameter $\mu(X_{e})=\log[h(N_{e};\betavec)]$ and shape parameter
$\sigma_{X}$. The logarithm of the $p$ quantile of $X$ is
$\log[x_{p}(N_{e})]=\mu(N_{e})+\Phi^{-1}(p)\sigma_{X} $. 

The $100(1-\alpha)\%$
LR confidence bands for the fatigue-strength qf $x_{p}(N_{e})$ for given $N_{e}$ over a
range of values of $p$
using~(\ref{equation:xi.maxL})) are
$\left[\,\xplower(N_{e}),\quad\xpupper(N_{e}) \,\right]$
where
\begin{align*}
\xplower(N_{e})&=\min\big\{ x_{p}(N_{e},\thetavec) \mid \loglike(\thetavec) = k
\big\}\\ 
\xpupper(N_{e})&=\max\big\{ x_{p}(N_{e},\thetavec) \mid \loglike(\thetavec) = k
\big\},
\end{align*}
$k=\loglike(\thetavechat )-(1/2)\chi_{(1-\alpha; 1)}^2$, and
$\thetavechat = (\betavechat,\sigmahat_{N})$.
Similarly, LR confidence bands for
the fatigue-strength cdf $F_{X}(x; N_{e})$
for a given $N_{e}$ and over a range of  $x$ values are
$[\,\, \Flower_{X}(x; N_{e}),\quad \Fupper_{X}(x; N_{e}) \,\,]$,
where
\begin{align*}
\Flower_{X}(x; N_{e})&=\min\big\{F_{X}(x; N_{e},\thetavec) \mid \loglike(\thetavec) = k
\big\}\\ 
\Fupper_{X}(x; N_{e})&=\max\big\{F_{X}(x; N_{e},\thetavec) \mid \loglike(\thetavec) = k
\big\}.
\end{align*}

\subsection{Likelihood Confidence Bands for the Induced Fatigue-Life
  Distribution}
\label{section:likelihood.conf.bands.induced.fatigue.life.distribution}

The fatigue-life random variable $N$ is
$N = h^{-1}[\exp(\log(S_{e})-\sigma_{X}\epsilon); \betavec].$
The cdf of fatigue life $N$ is
\begin{align*}
F_{N}(t; S_{e}) & =\Pr(N\leq t)=\Pr(\log[h(N;\betavec)]\geq
\log[h(t;\betavec)])\\
&=\Phi\left[\frac{\log(S_{e})-\log[h(t;\betavec)]}{\sigma_{X}}\right].
\end{align*}
The qf for $N$ is
$t_{p}(S_{e})=h^{-1}[\exp(\log(S_{e})-\Phi^{-1}(p)\sigma_{X});\betavec].$

LR confidence bands for the fatigue-life qf
at a given stress level $S_{e}$ and over a range
of values of $p$ are
$t_{p}(S_{e})$  using~(\ref{equation:xi.maxL})~is
$\left[\,\tplower(S_{e}),\quad\tpupper(S_{e})\,\right]$
where
\begin{align*}
\tplower(S_{e})
&=\min\big\{ t_{p}(S_{e},\thetavec) \mid \loglike(\thetavec) = k
                      \big\}\\
\tpupper(S_{e})
&=\max\big\{ t_{p}(S_{e},\thetavec) \mid \loglike(\thetavec) = k
                      \big\}.
\end{align*}
Similarly, LR confidence bands for the fatigue-life cdf
$F_{N}(t; S_{e})$ for a given level of stress $S_{e}$ over a range
 of $t$ values are
\begin{align*}
[\,\, \Flower_{N}(t; S_{e}),\quad \Fupper_{N}(t; S_{e}) \,\,]
\end{align*}
where
\begin{align*}
\Flower_{N}(t; S_{e}) &=\min\left\{F_{N}(t; S_{e},\thetavec) \mid
	\loglike(\thetavec) = k \right\}\\ 
\Fupper_{N}(t; S_{e}) &=\max\left\{F_{N}(t; S_{e},\thetavec) \mid
	\loglike(\thetavec) = k \right\}.
\end{align*}

\subsection{Equivalence Results When  Using a Specified
  Fatigue-Strength Model}\label{sec:equivalence.results}
There are similar equivalence results for the
specify-the-fatigue-strength-model approach that parallel the
results given in
Section~\ref{section:equivalence.results.specified.fatigue.life.model}. This
section states these explicitly. Proofs are omitted because they are
similar to those in
Section~\ref{section:equivalence.results.specified.fatigue.life.model}.
In particular, when the fatigue-strength model is specified and the
fatigue-life model is induced according to the approach described in
Section~\ref{section:induced.fatigue.life.model.and.ml}, we have the
following results:
\begin{itemize}
\item[\textbf{Result 4}] For any given value of number of cycles
  $N_{e}$, the  pointwise LR confidence bands for the
  fatigue-strength cdf and the corresponding fatigue-strength qf are
  equivalent.
\item[\textbf{Result 5}]
For any given level of stress $S_{e}$, pointwise LR
confidence bands for the induced fatigue-life cdf and the
induced fatigue-life qf are equivalent.
\item[\textbf{Result 6}]
For a fixed $p$, confidence bands for the fatigue-strength qf
for a range of number of cycles $N_{e}$ described in
Section~\ref{section:likelihood.conf.bands.specified.fatigue.strength.distribution}
and the induced fatigue-life qf
for a range of stress $S_{e}$ described in
Section~\ref{section:likelihood.conf.bands.induced.fatigue.life.distribution}
are equivalent.

\end{itemize}

\section{Estimating the Fatigue-Life and Fatigue-Strength Distributions
  of Nitinol Wire}
\label{section:estimating.quantiles.fatigue.life.nitinol.wire}
An experiment was conducted to study the fatigue life of nitinol
wire. Nitinol is a superelastic alloy that is used, for example, in
medical devices such as stents and artificial heart valves. The data
for this example were previously analyzed in~\citet{Falk2019}
and~\citet{Meekeretal2022} and are a subset of a larger data set
analyzed in \citet{Weaver_etal2022}.  The test consisted of 66 wire
specimens allocated across 10 levels of Percent Strain. The
specimens were subjected to rotary bend fatigue tests
under strain control until failure or until one billion cycles,
whichever came first. By the end of the test, there were 46 failures
and 20 runouts. The purpose of the test was to estimate small
quantiles of the fatigue-strength distribution at 600 million
cycles. Details of the estimation results presented in the cited
papers will not be repeated here.

\begin{figure}
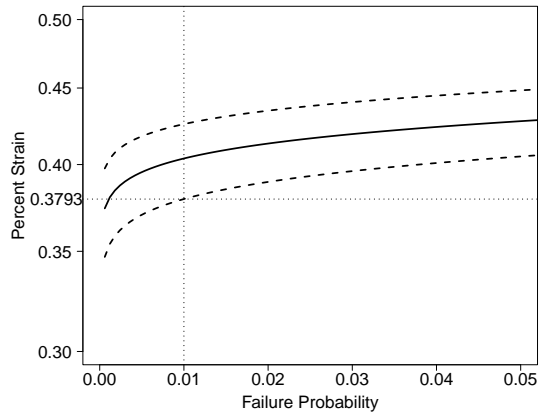

\begin{tabular}{cc}
\phantom{x}Fatigue-life failure probability estimates~(a) & \phantom{xxxxx}Fatigue-strength quantile estimates~(b) \\[-2ex]
\rsplidapdffiguresize{pdfFigures/Figure3a_Nitinol02StrengthDistLikelihoodIntervals}{3.1in}&
\rsplidapdffiguresize{pdfFigures/Figure3b_Nitinol02StengthQuantileLikelihoodIntervals}{3.1in}\\
\end{tabular}
\caption{Fraction failing as a function of percent strain for time
  600 million cycles~(a); Strength quantile as a function of percent
  strain for time 600 million cycles~(b); the dashed lines are pointwise 90\%
  confidence bands.}
\label{figure:nitinol.fatigue.strength.plots}
\end{figure}

Figure~\ref{figure:nitinol.fatigue.strength.plots}a is a plot of the
ML estimates of the fatigue-life
failure probabilities and corresponding bands of
pointwise LR two-sided 90\% confidence intervals, as a
function of Percent Strain, for a fixed value of $N_{e}$= 600
million cycles.
Plotting an ML estimate of $F_{N}(t; S_{e})$ (and the corresponding
confidence bands) versus
$S_{e}$ in this way (for fixed $t=600$ million cycles)
is also an ML estimate (and corresponding confidence bands) 
for the fatigue-strength cdf.

To obtain a lower confidence bound on the 0.01 quantile of
the nitinol fatigue-strength distribution at 600 million cycles,
we can use Result~4 in
Section~\ref{sec:equivalence.results}. In particular, the
needed blower bound can be obtained from the value of
Percent Strain that results in an one-sided upper 95\% confidence
bound on the fatigue-life failure probability
at 600 million cycles that is equal to 0.01
(i.e., $\tildex{F}_{N}(600; \textrm{Percent Strain})=0.01$).
Figure~\ref{figure:nitinol.fatigue.strength.plots}a
shows this value (note where the dotted lines cross) to be
$\undertilde{x}_{0.01}(600) = 0.3793$ Percent Strain.
Similar to the nib spring example in
Section~\ref{section:estimating.quantiles.fatigue.strength.distribution.new.spring},
such fatigue-life inferences can be obtained from commonly available
statistical software and then used to make fatigue-strength inferences.

Figure~\ref{figure:nitinol.fatigue.strength.plots}b focuses directly
on the needed inference, providing the ML estimate and 2-sided
pointwise 90\% confidence bands for the quantile function of the
fatigue-strength distribution. The figure shows
(note where the dotted lines cross) that the one-sided
lower 95\% confidence bound for the 0.01 quantile is again 0.3793
Percent Strain (i.e., $\undertilde{x}_{0.01}(600) = 0.3793$).

\section{Conclusions and Areas for Future Research}
\label{section:conclusions}
In this paper, we showed the equivalence of different sets of
confidence bands when those bands are based on inverting likelihood
ratio tests. These results are useful for making inferences on
fatigue-strength distributions by using readily available software
for estimation of fatigue-life distributions. The results will also
be useful for those who want to develop statistical software to
estimate fatigue-strength distributions.

When the confidence bands are based on the more commonly used Wald
method, the results are only approximate, with the approximation
improving with larger sample sizes. The intuition for this is
that Wald confidence intervals are based on a quadratic
approximation to the profile likelihood function that determines
LR confidence intervals,
as described in \citet{MeekerEscobar1995}.
It would be interesting to study the adequacy of this approximation
to help practitioners decide when it is appropriate to use
Wald-based intervals.



\bibliographystyle{chicago}
\addcontentsline{toc}{section}{\protect\numberline{}References}
\bibliography{ms}

\end{document}